\documentclass[iop]{emulateapj}

\usepackage{amsmath}
\usepackage{subfigure}

\def\slac{KIPAC, SLAC National Accelerator Laboratory, 2575 Sand Hill Rd, Menlo Park, CA 94025, USA}
\def\ukzn{Astrophysics and Cosmology Research Unit, University of KwaZulu-Natal,
Westville, Durban 4000, South Africa}
\def\llnl{Lawrence Livermore National Laboratory, 7000 East Ave, Livermore, CA 94550, USA}

\begin{document}

\title{

  {\large \bf \boldmath
Masked areas in shear peak statistics: a forward modeling approach
}
}
\author{D. Bard\altaffilmark{1}, J. M. Kratochvil\altaffilmark{2}, W. Dawson\altaffilmark{3}}
\email{djbard@slac.stanford.edu}
\altaffiltext{1}{\slac}
\altaffiltext{2}{\ukzn}
\altaffiltext{3}{\llnl}

\begin{abstract}
\noindent The statistics of shear peaks have been shown to provide valuable cosmological information beyond the power spectrum, and will be an important constraint of models of cosmology with the large survey areas provided by forthcoming astronomical surveys. 
Surveys include masked areas due to bright stars, bad pixels etc, which must be accounted for in producing constraints on cosmology from shear maps. 
We advocate a forward-modeling approach, where the impact of masking (and other survey artifacts) are accounted for in the theoretical prediction of cosmological parameters, rather than removed from survey data. 
We use masks based on the Deep Lens Survey, and explore the impact of up to 37\%\ of the survey area being masked on LSST and DES-scale surveys. 
By reconstructing maps of aperture mass, the masking effect is smoothed out, resulting in up to 14\%\ smaller statistical uncertainties compared to simply reducing the survey area by the masked area. 
We show that, even in the presence of large survey masks, the bias in cosmological parameter estimation produced in the forward-modeling process is $\approx 1\%$, dominated by bias caused by limited simulation volume. 
We also explore how this potential bias scales with survey area and find that small survey areas are more significantly impacted by the differences in cosmological structure in the data and simulated volumes, due to cosmic variance.

\end{abstract}

\maketitle

\section{Introduction}
\label{sec:intro}

Gravitational lensing is the phenomenon whereby light traveling through space is deflected by gravitational potentials in its path. 
This results in the distortion of the observed shapes of galaxies, and this distortion (called `shear') can be used to map the concentration of matter in the universe. 
The distortions of individual galaxies are small and  we have no knowledge of the original shape of the galaxies, so we must apply statistical analysis methods to obtain cosmological information from the ensemble of large numbers of galaxy shape distortions. 
In the past two years, measuring the statistics of the reconstructed shear field (shape distortion field) has gained ground as a competitive method of extracting cosmological information from weak lensing surveys. 
These statistics of shear peaks contain information beyond the power spectrum, and have been shown to be a potentially useful complement to the well-established measurement of the shear-shear correlation function.  

Counting peaks in shear maps was first proposed as a probe of cosmology by~\citet{Jain++2000}, who traced the dependence of shear peak counts with the fractional matter density of the Universe, $\Omega_m$. 
Early work emphasized that high-significance peaks in shear maps correspond to galaxy clusters, which have been long known to be a valuable probe of cosmology. 
However, this method of detecting clusters is prone to contamination from superpositions of structure along the line of sight and does not give a pure sample of clusters. 
In addition, voids along the line of sight can produce an under-estimate of cluster mass. 
As such, peaks in shear maps are not very useful as a method of identifying and measuring the mass of clusters specifically \citep[see, for example, ][]{Schirmer++2007, Dietrich++2007}, but nonetheless contain valuable cosmological information. 
\citet{Kratochvil++2010} and \citet{Yang++2011}, working with maps of convergence (confirmed by \citet{Bard++2013} working with maps of aperture mass) showed that in fact the majority of cosmological information from shear peak counts comes from low-significance peaks below a signal-to-noise ratio SNR$<$3. 
Despite the large number of spurious peaks due purely to noise in this SNR region, the large total number of low-significance peaks from small dark matter halos and filamentary structure provides significant constraining power on cosmological parameters. 

Beyond simply counting peaks in simulated shear maps, \citet{Kratochvil++2012}, \citet{Petri++2013} and \citet{Shirasaki++2013} have used Minkowski functionals to extract  more information from reconstructed convergence maps produced from cosmological N-body simulations. 
In a similar vein, \citet{Marian++2013} introduced statistics beyond the abundance to further constrain cosmology, including peak-peak correlation functions. 
Recent work has also demonstrated the advantage of combining information from shear maps with other lensing statistics. 
\citet{Hilbert++2012} combined measurements of the shear-shear correlation function, shear peak counts and lensing magnification (all from simulations) to demonstrate that the combination of all probes significantly constrains models of primordial non-Gaussianities beyond the correlation function alone. 
Encouragingly, \citet{Liu++2013} have studied the impact of magnification bias on shear peak counts in simulated convergence maps, and found that the combination of power spectrum and shear peak counts can help mitigate the impact of magnification and size bias on these statistics. 

The statistics of shear/convergence maps is now well established as a valuable probe of cosmology, complementary to the shear correlation function. 
So far work has focused on simulation studies, but attention is now turning to characterizing and mitigating the issues found in real data that will impact this type of statistic, for example the impact of measurement errors~\citet{Bard++2013} and spurious shear signals~\citet{Petri++2014}. 
Real survey data has areas unusable for weak lensing measurements, because  galaxies are partially or entirely obscured by stars, bad pixels or areas affected by blooming and bleed trails from saturated bright stars. 
Accounting for these masked areas is of particular importance in measuring any map statistic. 
Methods to deal with masked areas in the maps of shear or convergence have been proposed in the literature. 
\citet{VanderPlas++2012} have advocated an in-painting technique using Karhunen-Loeve analysis, where the information lost in the masked areas is reconstructed using the general statistical properties of the un-masked areas. 
These methods successfully recover the information lost to the masked areas, and allow convergence maps to be reconstructed free of the ``ringing'' effect produced when performing a Fourier transform of gappy data. 
However, they make assumptions about the underlying cosmology of the universe in order to fill in the missing data. 
This can introduce a bias on cosmological parameter estimation that is difficult to quantify. 

We count peaks as a function of SNR to account for noise in our reconstructed maps of convergence or aperture mass (see Section~\ref{sec:formalism} for details). 
Masks reduce the number of available galaxies and therefore distort the SNR around masked areas. 
The impact of this SNR distortion on Minkovski functionals calculated from simulated convergence fields was demonstrated by~\citet{Shirasaki++2013}. 
\citet{Liu++2014} evaluated this effect on shear peak counts of convergence maps, and proposed a way to correct for this effect on SNR by breaking maps into areas near and far from masks, and treating their statistics independently, based on the formalism introduced in~\citet{Fan++2010}. 

For convenience, nearly all previous work on the statistics of shear peaks has concentrated on maps of convergence $\kappa$. 
Convergence maps can be directly produced from ray-traced N-body simulations, but are hard to reconstruct from bounded, gappy observations due to the Fourier transforms required (although see~\citet{Fan++2010} for a method that avoids Fourier transforms). 
This paper addresses the issue of masked areas in maps of reconstructed aperture mass. 
The aperture mass is easily reconstructed from observational data, but cannot be directly produced by cosmological simulations. 
Aperture mass is calculated from a weighted sum over all the galaxies within a given aperture. 
The radius of this aperture is usually matched to the typical angle on the sky that is subtended by a galaxy cluster, a few arcminutes. 
This is much larger than the typical mask area in survey data. 
The loss of information in masking, and the resulting bias introduced in shear peak counting, comes from higher noise in the masked areas. 
The SNR of peaks in masked maps is systematically shifted downwards compared to peaks in unmasked maps. 
The calculation of aperture mass can smooth over this bias simply by averaging over many more galaxies than are missing in the mask.

In this work, we wish to emphasize the value of a forward-modeling approach to the statistics of shear peak counts. 
\citet{Maturi++2011} has made an analytic prediction of shear peak counts using Gaussian random fields, which met with success in the limit of high-significance peaks. 
Work since then \citep{Kratochvil++2010, Bard++2013} has shown that the majority of cosmological information exists in the medium and low peak regimes, and not only in the high peaks. 
In order to use shear peak counts to constrain models of cosmology, rather than make analytical predictions we must forward-model our predicted measurements. 
This involves creating mock catalogues of galaxy measurements, with data gaps, based on cosmological N-body simulations. 
Data can then be compared to the predictions made using the mock catalogues to find a best-fit to the cosmological parameters under consideration.

We measure shear using galaxies, which are noisy tracers of the underlying shear field because they have intrinsic shape. 
In forward-modeling our predictions, we must therefore include this imperfection, and also account for any errors that may by introduced by instrumental and measurement effects. 
Our previous work \citep{Bard++2013} created a framework for including galaxy shape noise and the measurement error that can be expected in a ten-year stack of LSST image data. 
We found that measurement error has a very small impact compared to shape noise, which is very encouraging for the robustness of this technique. 
In using a forward-modeling technique, we need to take into account all survey artifacts, including masked areas. 
When working with data from a survey telescope, we know the mask that has been applied to the data, and we can simply apply that same mask to our mock galaxy catalogues. 
If we are able to simulate the exact structure of our Universe, then we will have removed any bias in our measurement due to masking. 
In reality of course we cannot simulate the observed sky perfectly, but averaged over a large area the impact of differing structure between simulations and data will be mitigated. 
We investigate in this work the limit in which this holds.

This paper will apply the forward-modeling pipeline developed in \citet{Bard++2013} and include masking for the first time. 
We do not attempt to fill in the masked areas. 
Instead, we simply ignore them when we reconstruct our maps of aperture mass, and predict the shear peak counts from different cosmologies in the presence of masked areas. 
We consider the impact of masking on shear peak counts in maps of aperture mass and on the resulting constraints on cosmological parameters. 
The stellar density varies over the sky, so we examine the impact of large amounts of masking on the statistics of shear peaks.

In Section~\ref{sec:formalism} we cover the weak gravitational lensing formalism and introduce the aperture mass calculation. 
In Section~\ref{sec:masks} we describe the masks we use in this work, based on those used in the Deep Lens Survey~\citep{Wittman++2002}. 
We describe our forward-modeling approach in Section~\ref{sec:method}, using a combination of cosmological simulations and image simulations to produce mock galaxy catalogues with realistic shape noise and measurement error. 
The issues associated with limited simulation and survey volumes, and the impact of masking when applied to these volumes, is evaluated in Section~\ref{sec:results}. 
Finally, we conclude in Section~\ref{sec:conclusion}. 

\section{Formalism}
\label{sec:formalism}

Gravitational lensing deflects light emitted from distant galaxies, as it passes through the universe to our telescope through the gravitational field of all the matter along our line-of-sight. 
To measure this distortion, the observable we use in our surveys the ellipticity of the galaxies. 
This is a complex parameter $\epsilon = \epsilon_1 + i\epsilon_2$, where the components $\epsilon_1$ and $\epsilon_2$ are the normalized moments of the intensity of the light of the galaxy $I_{i,j}$ weighted by a Gaussian function $W(x_1,x_2)$:
\begin{eqnarray}
\epsilon_1 = \frac{I_{11} - I_{22}}{I_{11} + I_{22}}, \hspace{10pt} \epsilon_2 = \frac{2I_{12}}{I_{11} + I_{22}}, \hspace{50pt} \\
I_{ij} = \frac{ \int \int W(x_1, x_2) f(x_1, x_2) x_i x_j dx_1 dx_2} {\int \int W(x_1, x_2) f(x_1, x_2) dx_1 dx_2}, \hspace{10pt} i, j = 1,2. 
\end{eqnarray}
\citep[This is the quantity $\chi$, or `distortion', as defined in][]{Schneider2005}. 
In this section we will address how the observed quantity $\epsilon$ can be used to obtain information about the matter in the universe, and how we construct the aperture mass from our measurements.

Following the formalism introduced in \citet{BernsteinJarvis2007}, the distortion of galaxy shapes by a gravitational lensing potential is described by the Jacobian
\begin{equation}
A = (1 - \kappa) \left( \begin{array}{cc} 1-g_1 & -g_2 \\ -g_2 & 1+g_1 \end{array} \right). 
\end{equation}
$g = \frac{\gamma}{1-\kappa} $ is the observable reduced shear constructed from the complex parameters describing the effect of lensing on observed galaxy shape  $\gamma \equiv \gamma_1 + i\gamma_2$, and the magnification of galaxy image $\kappa$, the convergence. Note that in the weak lensing regime, for small shears applied to round objects, $\gamma \approx g \approx \frac{\epsilon}{2}$

Galaxies have intrinsic shape and random orientations on the sky, so the shape measured in surveys is a combination of the galaxy shape and the distortion due to gravitational lensing. 
The uncertainty in a measurement of $g$ is $\sigma_g$, and is a combination of both galaxy shape noise $\sigma_{int}$ and measurement error $\sigma_{meas}$, $\sigma^2_g = \sigma^2_{int} + \sigma^2_{meas}$, where ``error'' is defined as the difference between the measured and true quantity, and ``uncertainty'' as the standard deviation of the difference between the measured and true quantities. 

A matter over-density along the line-of-sight will cause the observed shape of  background galaxies to be tangentially aligned around the projected mass peak. 
We can use this fact to detect mass peaks by constructing a weighted sum over the tangential components of galaxy shapes around a point. 
This quantity is called the aperture mass, and is defined in \cite{Schneider2005} as:
\begin{equation}
M_{ap}(\theta_0) = \int d^2 \boldsymbol{\theta} Q(\theta) g_t(\theta, \theta_0),
\end{equation}
where $Q$ is a weighting function and $g_t$ is the tangential component of reduced shear relative to $\theta_0$ defined as 
\begin{equation}
g_{\mathrm{t}}(\theta, \theta_0)=-(g_1\cos(2\phi)+g_2\sin(2\phi)).
\end{equation}
$\phi$  is the angle with respect to the horizontal axis between positions $\theta_0$ and $\theta$ in the map. 
In practice, we do not measure the shear field directly but use observed galaxies to sample it, so that the aperture mass becomes a sum over the tangential components of galaxy shapes:
\begin{equation}
M_{ap}(\theta_0) = \frac{1}{N_g} \sum^{N_g}_{i=1}  Q(\theta) g_{i,t}, 
\label{eqn:map}
\end{equation}
where $N_g$ is the number of galaxy images within the aperture. 
If the weight function $Q$ follows the expected shear profile of a mass peak then the aperture mass is a matched filter for detecting mass peaks. 
We use the spherically symmetric function introduced by ~\citet{Schirmer++2007}, which follows an NFW~\citep{Navarro++1996} profile with exponential cutoffs as $x \to 0$ and $x \to \infty$, where $x=\theta_i/\theta_{max}$, where $\theta_{max}$ gives the radius to which the filter is tuned:
\begin{equation}
Q_{NFW}(x, x_c) = \frac{1}{ 1 + e^{6-160x} + e^{-47+50x}} \frac{\tanh(x/x_c)}{x/x_c}. 
\end{equation}
$x_c$ is a constant, set to 0.15, which has been empirically determined to be a good value for shear peak counting~\citep{Hetterscheidt++2005}.
In previous work, in common with \citet{DietrichHartlap2010} and \citet{Marian++2013}, we used a value of 5.6\arcmin.

We must consider the effect of noise (both shape noise and measurement error) in our maps, so we construct maps of SNR from our mock galaxy catalogues. 
As described in \citet{BS2001}, in the absence of lensing the rms dispersion of $M_{ap}$ is determined from the shape noise of the galaxies:
\begin{equation}
\sigma_{m_{ap}} = \frac{\sigma_g}{\sqrt{2}N_g} \sqrt{\sum_i Q^2(\theta_i)}. 
\end{equation}
Provided we are in the regime of weak lensing, $\sigma_{m_{ap}}$ will be close to the rms dispersion in the presence of lensing, and can be used to measure the uncertainty of the aperture mass directly from the data. 
We can therefore define the SNR at a point $\theta_0$ to be 
\begin{equation}
\mathrm{SNR}(\theta_0) = \frac{\sqrt{2} \sum_i Q(\theta_i)g_{i,t}} {\sqrt{\sum_i Q^2(\theta_i)g_i^2}}. 
\end{equation}

The statistic we use to constrain cosmology is the number of peaks in our SNR maps, where we progressively raise the SNR threshold. 
We define a peak as a group of pixels above a SNR threshold that have 8-connectivity - that is, that are connected along the sides or by the corners. 
This is the same definition used in \citet{DietrichHartlap2010} and \citet{Bard++2013}. 

\section{Masks}
\label{sec:masks}

An all-southern-sky survey like LSST will encounter new issues related to sky coverage that are not an issue for smaller survey volumes. 
For example, at low galactic latitude nearly 100\%\ of the sky will be masked due to bright stars and precise measurements of galaxies will not be possible. 
In addition, reddening from dust around the galactic center would also impact the quality of galaxy measurements, impacting photometric redshift measurements. 
A full exploration of how these things impact the effective area of a survey is a topic for another paper; here we simply consider the effect of masked areas due to stellar density. 
We wish to determine how masking affects measurements of the statistics of shear peaks from aperture mass measurements. 
For this, we take masks used by the Deep Lens Survey.

The Deep Lens Survey \citep{Wittman++2002} is a deep multi-band imaging survey of five 4 sq. degree fields with two 4-meter telescopes at Kitt Peak and Cerro Tololo.
DLS is the deepest optical survey to date among the current $>10$ sq. degree surveys, reaching a mean source redshift of $z=1$ and a limiting r-band magnitude of 26. 
Regions affected by bright stars (such as PSF wings, bleeding and diffraction spikes) were masked out in the survey. 
Bad pixels were also masked out. 
The DLS fields are representative of a deep survey similar to LSST, and we chose to use the DLS masks as examples of an LSST mask might look like. 

The five DLS fields have different levels of masking, ranging from 6.1\%\ -13.4\%, given in Table~\ref{tab:DLS-masks}.  
The DLS fields are 4 square degrees, which is significantly smaller than our lensing maps of 12 square degrees (see the next section for details). 
To adapt a DLS mask to one of our lensing maps, it is tiled, rotated and shifted. 
Each lensing map has a unique mask applied. 
From these masks, we create masks for higher stellar density by offsetting and layering different combinations of the DLS masks. 
The densest mask we create has 37.6\%\ of the sky masked, which is the maximum masked we obtain when layering all five of the DLS masks. 

\begin{table}
\centering
\caption[]{\textit{DLS fields and masks}}
\begin{tabular}{|c|c|c|c| }
\hline
DLS field & Galactic longitude & Galactic latitude & Masked area \\ \hline
F1      &123:41:24.03	&-50:18:12.78 & 6.4\% \\
F2	&196:10:04.7	&+43:28:11.42 & 6.9\% \\ 
F3	&255:30:36.16	&-34:49:01.17 & 13.1\% \\
F4	&256:29:00.64	&+46:48:58.80 & 8.4\%\ \\ 
F5	&327:37:28.82	&+49:47:59.04 & 10.1\%\ \\ \hline
\end{tabular}
\label{tab:DLS-masks}
\centering
\end{table}

\begin{figure}
\includegraphics[width=3in]{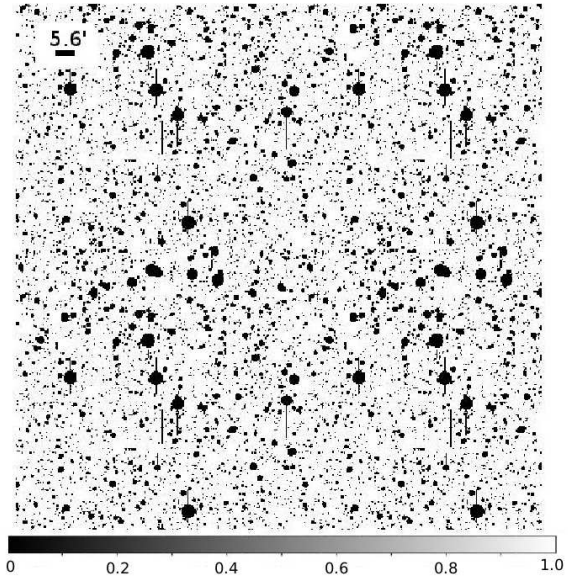}
  \caption{Example mask based on the DLS field F3, covering 4 square degrees. The mask is tiled so that it fits the simulated shear maps of 12 square degrees, and is randomly rotated for each shear map. 13.4\%\ of the sky has been masked. }
\label{fig:F3mask}
\end{figure}

Table~\ref{tab:masks} gives the different levels of masking we use in this paper, and Figure~\ref{fig:F3mask} shows one example mask created from the mask for DLS F3, covering 13.4\%\ of the sky. 

\begin{table}
\centering
\caption[]{\textit{Composite masks used in this paper}}
\begin{tabular}{|c|c| }
\hline
Proportion of sky masked & Superposition of DLS masks\\ \hline
13.4\%  & F3\\
20.5\%  & F3+F4\\
28.4\%  & F3+F4+F5\\
37.6\%  & F1+F2+F3+F4+F5\\  \hline
\end{tabular}
\label{tab:masks}
\centering
\end{table}

\section{Method}
\label{sec:method}

Our forward-modeling approach combines shear maps from ray-traced cosmological N-body simulations and galaxies with realistic properties and shape measurement errors determined from the LSST Image Simulator \citep[ImSim][]{Peterson++prep}. 
This methodology has been described in detail in \citet{Bard++2013}, and we give here a brief summary. 

 \subsection{Cosmological Simulations}
We use a suite of cosmological N-body simulations described in \citet{Kratochvil++2010}, used in several previous papers on the statistics of shear peaks ~\citep{Yang++2011, Kratochvil++2012, Bard++2013}. 
The simulation consists of $512^2$ particle in a box, with a box size $240h^{-1}$Mpc, produced using a modified version of the publicly-available Gadget-2 N-body code \citep{Springer2005}. 
The linear matter power spectrum, used as input to the initial conditions generator, was produced using CAMB \citep{Lewis++2000}. 
The cosmological parameters varied in these simulations are the fractional matter density of the universe $\Omega_m$ and the normalization of the matter power spectrum at the length scale $8h^{-1}$Mpc $\sigma_8$. 
$w$, the parameter describing the evolution of the  equation-of-state of the universe, is held constant at $w=-1.0$. 
In this work we will therefore not be able to quantify the impact of masking on estimation of $w$; we leave this to future work. 
Six sets of simulations were produced - two for the fiducial cosmology, and four with variations on the cosmological parameters. 
The fiducial cosmology is chosen to have parameters {$\Omega_m$ = 0.26, $\Omega_{\Lambda}$=0.74, $w$=-1.0, $n_s$ = 0.96, $\sigma_8$ = 0.798, $H_0$ = 0.72}. 
These are close to the current best-fit parameters of the universe as determined by the Planck survey \citep{Planck}. 
The variations of parameters in our simulations is given in Table~\ref{tab:cosmos}.

Five independent runs of each simulation were produced, using the same input power spectrum but producing a statistically robust sets of simulations.  
By randomly rotating and shifting the simulation data cubes, we obtain 1000 quasi-independent lines of sight, which are then ray-traced to produce maps of shear and convergence.  
The fiducial cosmology has two sets of entirely independent maps, one produced from a set of 5 N-body runs and one from a set of 45 runs. 
This allows us to use one as an independent `data' set in our analysis, and the other to define the fiducial point in parameter space. 
The ``auxiliary'' map set, produced from the set of 5 runs, is generated using the same five quasi-identical initial conditions as the non-fiducial cosmologies. 
The ``primary'' map set is produced from the set of 45 N-body runs. 

The ray-tracing for a flat sky is performed using the algorithm described in \citet{Hamana++2004}. 
Shear and convergence maps are produced for 2048$\times$2048 light rays at three redshift planes $z = [1.0, 1.5, 2.0]$ for each of the simulations listed in Table~\ref{tab:cosmos}. 
Each map covers 12 square degrees. 
This gives us a total map set of 3000 maps per cosmology, and a grand total of 18,000 maps to be processed. 

\begin{table}
\centering
\caption[]{\textit{Cosmological parameters used in our simulations, their identifiers and the number of independent simulations produced for each parameter set.}}
\begin{tabular}{|l|c|c|c|c|c|} 
\hline
WL Map Set & $\sigma_8$ & $w$ & $\Omega_m$ & $\Omega_\Lambda$ & \# of \\
Identifier & & & & & sims \\
\hline
Primary & 0.798 & -1.0 & 0.26 & 0.74 &45\\
Auxiliary & 0.798 & -1.0 & 0.26 & 0.74 & 5\\
om23 & 0.798 & -1.0 & 0.23 & 0.77 & 5\\
om29 & 0.798 & -1.0 & 0.29 & 0.71& 5\\
si75 & 0.750 & -1.0 & 0.26 & 0.74 & 5\\
si85 & 0.850 & -1.0 & 0.26 & 0.74 & 5\\
\hline
\end{tabular}\label{tab:cosmos}
\centering
\end{table}

\subsection{Source Galaxies}
We need to account for both shape noise and measurement error in our galaxies. 
Since measurement error depends on the size and magnitude of the galaxy, we therefore need to draw a realistic distribution of galaxy properties. 
Many of these properties are taken from galaxy catalogues developed for use as input to the LSST Image Simulator. 
These catalogues consist of galaxies produced in semi-analytic models from the Millennium simulation, matched to a compilation of observations from deep survey data\footnote{http://astro.dur.ac.uk/\textasciitilde nm/pubhtml/counts/counts.html}, the DEEP2 survey~\citep{coil04}, and data from the publicly available Hubble Deep Field catalogues\footnote{http://www.stsci.edu/ftp/science/hdf/archive/v2.html}. 
A comprehensive validation of the ImSim input catalogue is given in \citet{Peterson++prep}. 

We place source galaxies at random locations across the 12 square degree lensing maps, to obtain a source density of 30 galaxies arcmin$^{-2}$ which is roughly the effective number of galaxies expected for LSST \citep{Chang++2013}. 
Placing galaxies at random entirely neglects the cosmological structure of the real galaxy distribution, and also neglects any systematic effects due to intrinsic alignments and magnification bias. 
Correlating galaxy location with shear maps without additional information from the initial N-body simulations is very difficult, and so we neglect these effects in this work. 
We draw an intrinsic ellipticity for each galaxy from a distribution based on measurement by the COSMOS survey \citep{Joachimi++2013, Leauthaud++2007} which found the width of the intrinsic shape noise distribution to be 0.23 per component of reduced shear. 
We note that the distribution of galaxy ellipticities is not Gaussian in shape, but more cuspy towards zero, and we account for this fact in our mock catalogue. 
Galaxy redshift, size and magnitude are all assigned based on quantities drawn from the ImSim input catalogues. 
The magnitude and size are redshift-dependent. 

Galaxy shapes are then sheared according to their location and redshift, using values from the simulated shear and convergence maps interpolated to the position of the galaxy in RA, dec and redshift. 
At this point we have a mock catalogue of perfectly measured galaxies for our simulated cosmology. 

\subsection{Assigning measurement errors}
We next assign a measurement error to each galaxy. 
These measurement errors have been determined from a large suite of image simulations performed using ImSim, where LSST-type observations of the sky are simulated and analyzed. 
We simulate the same area of the sky 100 times, each time with a different atmospheric realistically. 
The seeing is taken from a distribution produced by the LSST Operations Simulator~\citep{opsim} of median seeing 0.7\arcsec \citep[which is considered acceptable quality for weak lensing analysis in LSST,][]{LSST-WL}. 
We apply a constant shear to the simulated images, and repeat for a wide variety of shear values. 
The galaxies simulated in this set of image simulations are circular, since we wish to isolate the effect of measurement error from that of intrinsic shape noise. 
The simulated images are processed using SourceExtractor~\citep{sEx}, and the resulting catalogue of sources is separated into candidate stars and galaxies using a simple cut on object size and maximum surface brightness. 
The stars are used to determine the ellipticity components of the PSF, which is interpolated to the galaxy locations using a third order polynomial function. 
The KSB algorithm~\citep{ksb}, as implemented in the {\it{imcat}} pipeline, is then used to deconvolve the PSF from the galaxy shapes, giving us the reduced shear of each galaxy. 
We ``stack'' the measurements of the galaxies over the 100 atmospheric realizations by simply averaging their measured reduced shear. 
In order to obtain the error on the reduced shear measurement, we then average over all galaxies in our simulated field and compare the resulting reduced shear measurement for each (circular) galaxy to the reduced shear that we applied to the simulated images. 
This distribution is our shear measurement error. 
Note that we do not assign photometric errors, and that we assume perfect photo-z estimation.

As described in \citet{Bard++2013}, we find no dependence of measurement error on the input shear, nor on the measured shape of the galaxy. 
However, there is a correlation with the magnitude of the galaxy, where fainter galaxies have larger measurement errors. 
The measurement error that we apply to our set of mock galaxies is therefore magnitude-dependent. 

We create an independent mock galaxy catalogue for every one of our 1000 map sets, and we use the same galaxies (with different shear applied as appropriate) for the maps sets of each of the 6 simulated cosmologies. 
This galaxy catalogue is then masked. 
Galaxies in locations covered by our set of simulated masks are simply removed from the catalogue. 

The final step is to apply the aperture mass algorithm to the mock catalogues, and use the resulting map of SNR to count aperture mass peaks. 
We use a GPU implementation of the aperture mass algorithm, described in \citet{Bard++2013gpu} and used previously in~\citet{Bard++2013}. 
This speeds the calculation of the aperture mass by a factor of a few hundred, compared to a brute-force calculation on the CPU, and does not require any approximation. 
Since we have several thousand maps to process, this speed-up in calculation time is exceedingly useful. 
Finally, we have a set of 6,000 SNR maps - 1000 maps for each of 6 cosmologies. 

\begin{figure*}
\subfigure[ no mask]{\includegraphics[width=2.2in]{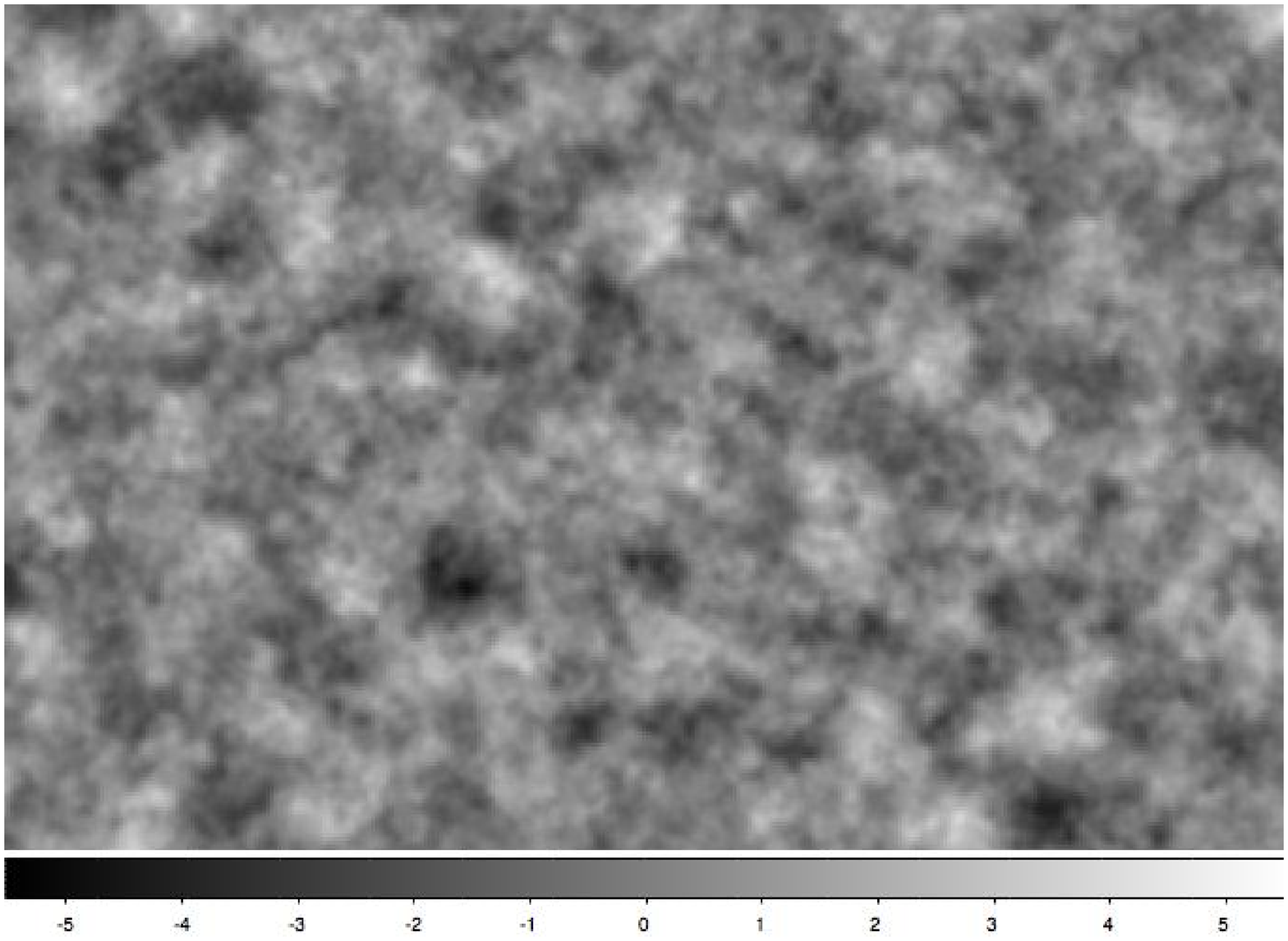}}
\subfigure[ 13.4\%\ mask]{\includegraphics[width=2.2in]{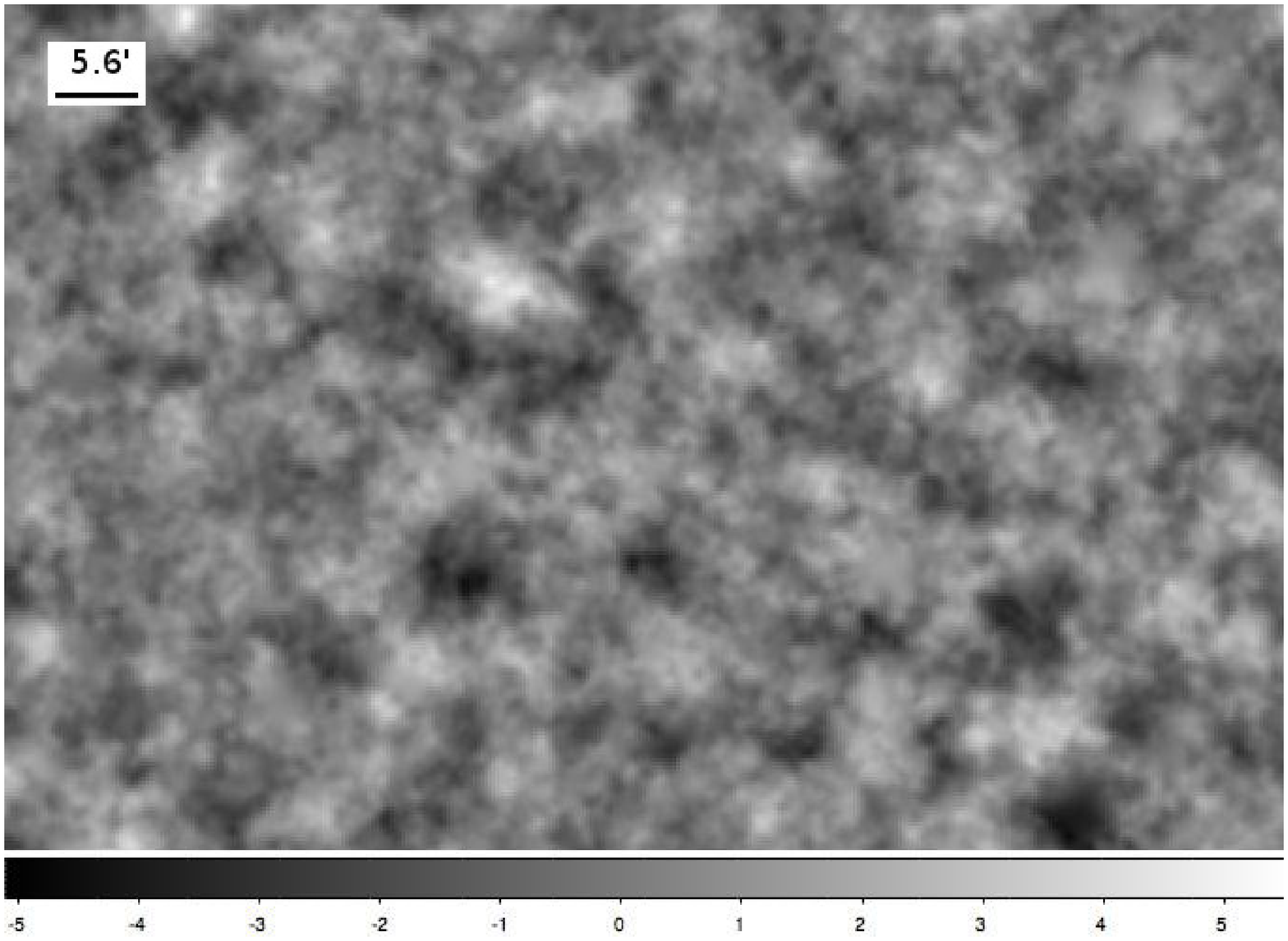}}
\subfigure[ 13.4\%\ mask (mask overlaid)]{\includegraphics[width=2.2in]{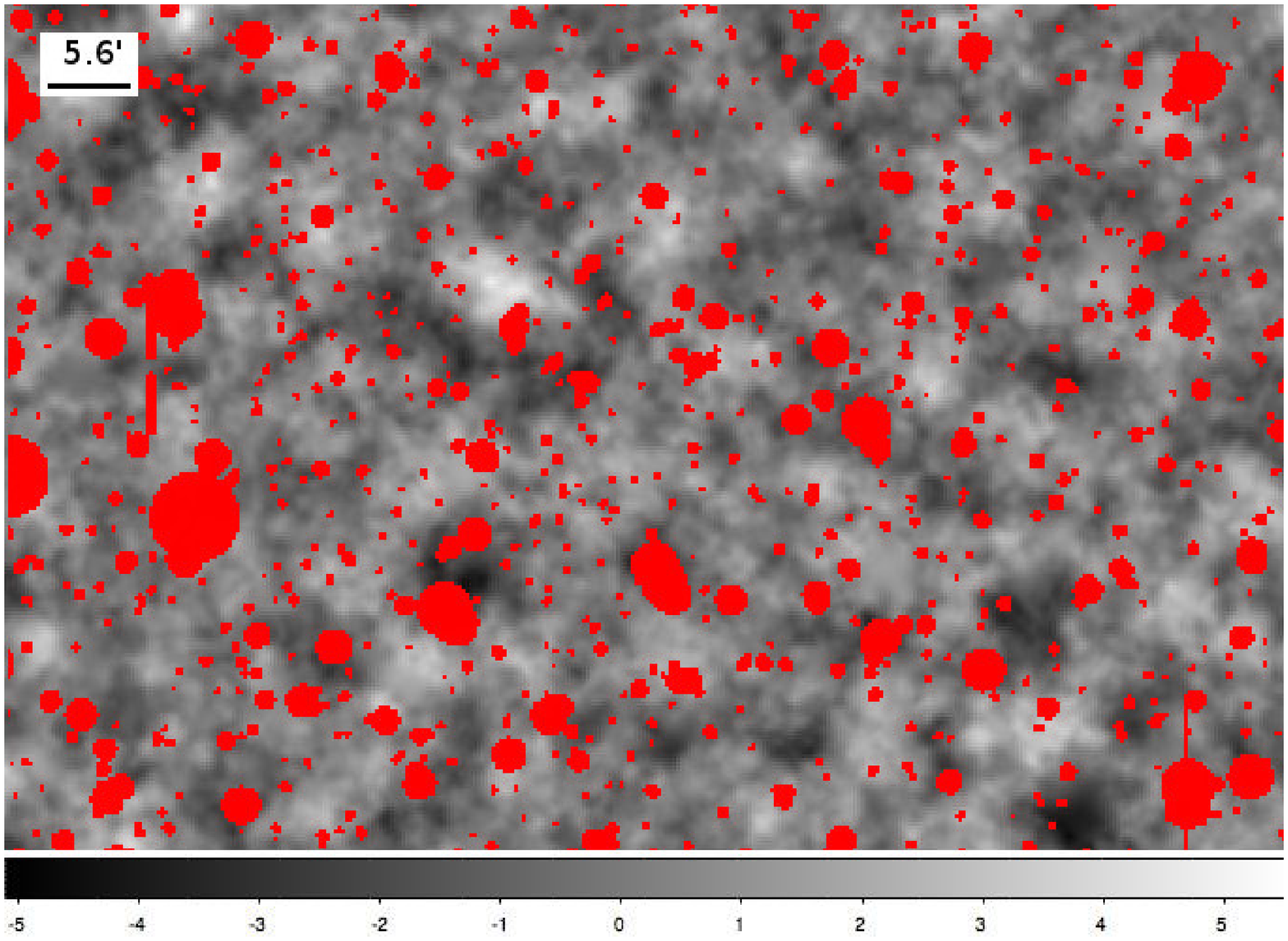}}
\caption{Sections of SNR maps with DLS field F3 masking applied, constructed using the aperture mass statistic. The radius of the smoothing kernel, 5.6', is shown for comparison, and is larger than the scale of most masked areas. }
\label{fig:masked-SNR}
\end{figure*}

We count peaks in our maps of SNR using the methodology described in \citet{DietrichHartlap2010}.
We apply a range of thresholds to each map, and count the number of peaks that appear above that threshold. 
Figure~\ref{fig:masked-SNR} shows a small area of one of the SNR maps for the fiducial cosmology with and without masking. 
There are anomalously smooth areas when masking is applied that will impact peak counts if not accounted for correctly. 

\subsection{Extracting Cosmological Parameters}
\label{sec:calc}
In order to quantify the impact of masking on cosmological parameter constraints, we calculate best-fit parameters using the methodology described in \citet{Bard++2013} and \citet{Liu++2013}. 

The aperture mass peak counts are histogrammed into 25 evenly-spaced bins in SNR. 
We calculate the average aperture mass peak counts $\overline{N_i}$ for SNR bin $i$ by averaging over 1000 maps from each simulated cosmology, in each SNR bin. 
This gives us the mean histogram of peaks counts in each SNR bin for the cosmologies listed in Table~\ref{tab:cosmos}. 
In order to extrapolate to areas of cosmological parameter space that we did not explicitly simulate, we perform a Taylor expansion around the fiducial parameter point to obtain the histogram of average peak counts, treating each bin in SNR independently:
\begin{equation}
\overline{N_i}({\bf{p}}) \approx \overline{N_i}({\bf{p_0}}) + \sum_{\alpha}{  \frac{\overline{N_i}(p^{(\alpha)}) - \overline{N_i}(p_0)} {p_{\alpha}^{(\alpha)} - p_{0,\alpha}}} . (p_{\alpha} - p_{0,\alpha}).
\end{equation}
This is the finite difference derivative, where $p_{\alpha}$ denotes one of the cosmological parameters ${\bf{p}} = (\sigma_8, \Omega_m)$, 
${\bf{p}^{(\alpha)}}$ is the cosmological parameter vector for a simulated non-fiducial cosmology, and ${\bf{p}}_0$ is the cosmological parameter vector for the fiducial cosmology. 
$\overline{N_i}({\bf{p_0}})$ is therefore the average number of peaks in SNR bin $i$ for the fiducial cosmology. 

We use an independent set of fiducial cosmology maps as our mock dataset. 
To find the best-fit cosmology of our `data', we follow the methodology used in \citet{Kratochvil++2012, Bard++2013, Liu++2013} and minimize the $\chi^2$ for the peak distribution of one `data' map using:
\begin{equation}
\chi^2({\bf{p}}) = \sum_{i,j} \Delta N_i({\bf{p}}) C^{-1}_{i,j} \Delta N_j({\bf{p}}).
\end{equation}
Here, $\Delta N_i({\bf{p}}) = N_i({\bf{p}}_0) - \overline{N_i}({\bf{p}})$ is the difference between the peak counts in a given map $N_i$ and the average peaks counts of the model $\overline{N_i}$ in the $i$th SNR bin. 
$C^{-1}_{i,j}$ is the inverse of the covariance matrix, estimated from the simulations where 
\begin{equation}
\label{eqn:covmat}
C_{ij}(\mathbf{p})\equiv\frac{1}{R-1}\sum_{r=1}^R [N_i(r,\mathbf{p})-\overline{N}_i(\mathbf{p})][N_j(r,\mathbf{p})-\overline{N}_j(\mathbf{p})].
\end{equation}
This covariance matrix contains contributions both from the sample variance of the true aperture mass signal and from the noise contributions.

As described in~\citet{Liu++2013}, if we make the assumption that peak counts depend linearly on the three cosmological parameters we have varied in our simulations, we can solve the $\chi^2$ minimization analytically. 
Using the notation given in~\citet{Liu++2013}, we define
\begin{eqnarray}
X_{i,\alpha} = \frac{ \partial \overline{N_i}} {\partial p_{\alpha}}\\
Y_i = N'_i - \overline{N_i}({\bf{p}}). 
\end{eqnarray}

Our expression for the $\chi^2$ therefore becomes:
\begin{eqnarray}
\Delta N_i = Y_i - X_{i,\alpha}dp_{\alpha}\\
\chi^2 = (Y_i - X_{i,\alpha}dp_{\alpha}) C^{-1}_{i,j} (Y_i - X_{i,\alpha}dp_{\alpha}). 
\end{eqnarray}

To minimize, we set $d\chi^2/d(dp_{\alpha}) = 0$ and obtain:
\begin{equation}
X_{i,\alpha} C^{-1}_{i,j} (Y_j - X_{j,\beta}dp_{\beta}) +  (Y_j - X_{j,\beta}dp_{\beta}) C^{-1}_{i,j} X_{j,\alpha} = 0.
\end{equation}

This is symmetric in $i$ and $j$, and so the two terms can be combined and the difference between the best fit and the fiducial cosmology can be written as
\begin{equation}
dp_{\beta} = (X_{i,\alpha} C^{-1}_{i,j} X_{j,\beta})^{-1} (X_{i,\alpha} C^{-1}_{i,j} Y_{j}). 
\end{equation}

Applying this fitting procedure to our 1000 maps, we have 1000 best-fit points.

There are limitations to our method, due to the (small) differences in cosmological structure between the primary and auxiliary fiducial simulations, and due to the limited sampling of cosmological parameter space. 
We have 1000 simulated maps in six cosmologies that we wish to use to emulate a cosmological measurement. 
To do this, we take the set of 1000 primary fiducial cosmology simulations to represent our `data' set, and use the auxiliary fiducial map set to define the fiducial point in parameter space. 
We use the average of the peak-count histograms for all 1000 maps of the auxiliary fiducial cosmology as the base of our Taylor expansion. 
This is the anchor of our parameter space. 
The `data' set and fiducial point simulation are therefore statistically independent, being drawn from the same underlying cosmology but independent realizations of it. 
This represents a somewhat optimistic scenario in which we have simulated the correct universe, neglecting baryonic effects. 
If we are able to fully characterize our observable (the histogram of lensing peak counts) across parameter space, then any of our simulations could act as the base of the Taylor expansion. 
It is important that the auxiliary fiducial simulation also be used to form the covariance matrix, since it is statistically independent from the simulations used as `data'. 
 This avoids correlations between the fitted data and the covariance matrix. 
It is also important that we use the map set with the largest statistical scatter (i.e. based on the largest number of simulations) as our `data' set, since this will determine the error on the bias. 

To test the validity of our assumption that our observable (the histogram of peak counts in SNR bins) varies linearly with cosmological parameter in our Taylor expansion, we used non-fiducial cosmologies as the anchor of the Taylor expansion, and found an additional bias on the order of a percent. 
The non-zero additional bias we see in using non-fiducial cosmologies as the anchor of the Taylor expansion reveals the limit of this assumption. 
To fully characterize the variation of our observable in the parameter space, even larger cosmological simulations are required. 

\section{Results}
\label{sec:results}

We have established how we produce masks and how we apply them to our mock galaxy catalogues. 
In this section we evaluate the impact that the masking has on cosmological constraints.

\subsection{The Impact of Masking on Peak Counts}

\begin{figure}
\includegraphics[width=3.2in]{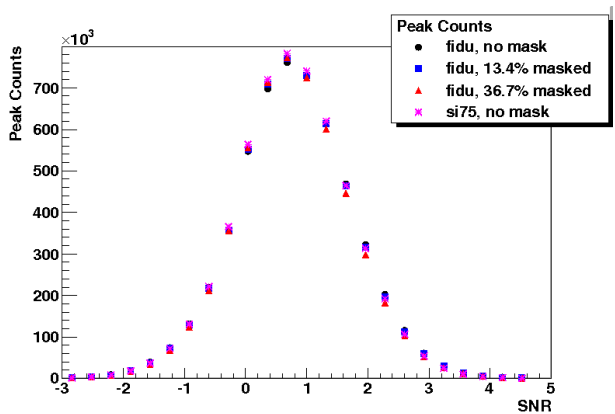}
\caption{Peak counts for the fiducial cosmology with different levels of masking, and for the $\sigma_8=0.75$ cosmology with no masking.}
\label{fig:peaks}
\end{figure}

We first consider the impact of masking on peak counts from aperture mass maps constructed using our nominal smoothing radius of 5.6\arcmin. 
Figure~\ref{fig:peaks} shows the peak counts at increasing SNR threshold for the fiducial cosmology, comparing the case of no mask, the DLS F3 mask with 13.4\%\ area covered, 37\%\ masking, and the peak counts for the $\sigma_8=0.75$ cosmology. 
Peaks at negative SNR come from voids or local under-densities along the line-of-sight. 

\begin{figure}
\includegraphics[width=3.2in]{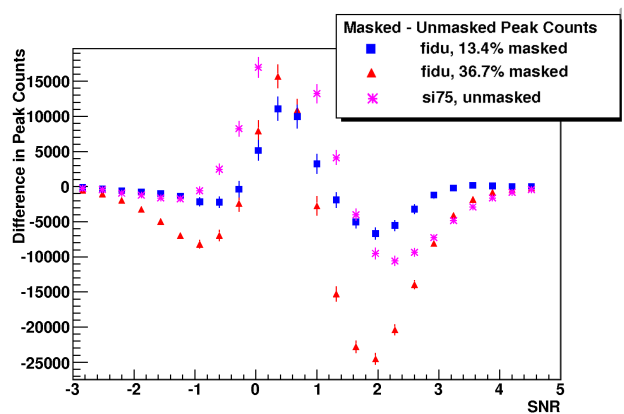}
\caption{Difference between peak counts for the fiducial cosmology and the same cosmology with different levels of masking, and the $\sigma_8=0.75$ cosmology with no masking.}
\label{fig:peaksdiffs}
\end{figure}

Figure~\ref{fig:peaksdiffs} shows the difference in peak counts between the masked and unmasked fiducial cosmology, and the unmasked si75 cosmology. 
As seen in previous work examining peaks in maps of convergence~\citep{Liu++2014}, adding masks decreases the number of peaks at low and high SNR, and slightly increases the number of peaks counted at mid-SNR (-0.5$<$SNR$<$ 1.5). 
This difference is significantly larger than the comparison to an alternative cosmology model, indicating the significance of masking for cosmological parameter estimation. 

\subsection{Impact of Masking on Cosmological Parameter Constraints}

We find that the average best-fit cosmological parameters calculated as described in Section~\ref{sec:calc} (with no masking applied) have a small (O(1\%)) bias. 
This is due to using the auxiliary cosmological simulations as the anchor of our parameter space in the Taylor expansion - if we use primary simulation mapset as both the anchor and the mock data, we have no bias. 
A 1\%\ bias is very small (and negligible for current astronomical datasets), but in the era of precision cosmology with LSST we expect to be able to measure cosmological parameters to 1\%\ statistical uncertainty. 
This bias reveals how much the small differences in structure between our two fiducial simulations matter, even averaged over 12,000 sq deg of simulated lensing maps. 
We plan future work with even larger simulation volumes to determine how much simulation volume is required before this bias is negligible. 
In the following, we correct for this 1\%\ bias, and consider the additional bias introduced due to masking effects.

In the following, we evaluate the impact of masking with respect to the small bias in parameter estimation we see using the auxiliary fiducial simulations as the base of the Taylor expansion.  
To scale to a survey the area of LSST, we need 1500 of our 12 square degree maps, so we scale our measurement by sampling (with replacement) 1500 maps from the 1000 we have available.

\begin{figure}[htb]
\includegraphics[width=3.5in]{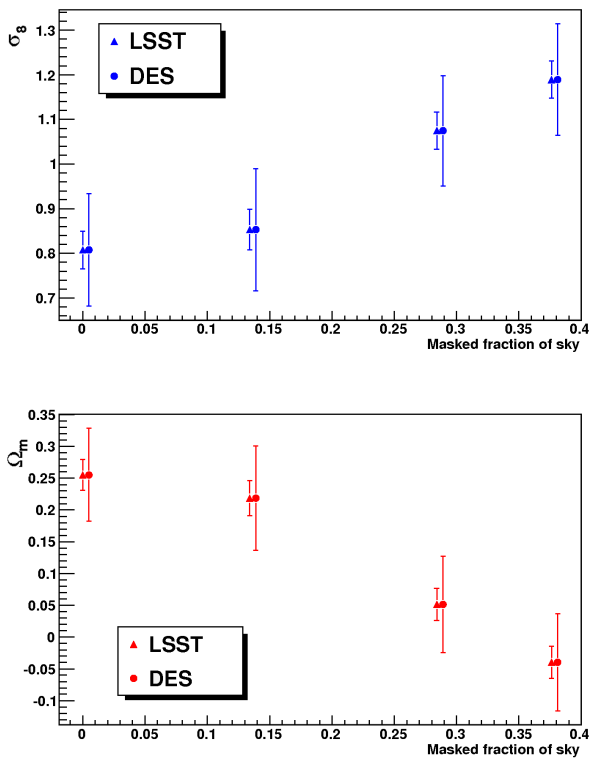} 
\caption{Cosmological parameter constraints obtained if a masked dataset is compared to an unmasked set of cosmological simulations, for LSST and DES survey volumes. Points are the mean best-fit parameter estimate (offset for clarity), and the errors shown are the standard errors on the mean multiplied by a factor of ten to make them visible. }
\label{fig:bias-unmasked}
\end{figure}

First, we consider the case where we neglect the impact of masking. 
In Figure~\ref{fig:bias-unmasked}, we show how our measurements of cosmological parameters are affected if the `data' mapset is masked, but the maps used in the Taylor expansion and covariance matrix are unmasked. 
A large bias in parameter estimation is produced, which grows  as the masked area grows. 
The parameter uncertainties expected for LSST (as calculated in this analysis) are on the order of 1\%, so the bias reaches tens of sigma. 
We have calculated expected uncertainties for a DES measurement by scaling our LSST uncertainties to account for the reduced the DES survey area (5000 square degrees) and expected galaxy density (12 per square arcminute\footnote{www.darkenergysurvey.org/reports/proposal-standalone.pdf}). 
These uncertainties are large enough that the bias produced from neglecting a 13\%\ masked area is less than 1$\sigma$ for both DES and LSST, but would be non-negligible beyond a 20\%\ masked area for LSST (and a 30\%\ masked area for DES). 
We note that this is only an estimate, as we have neglected the many other differences between LSST and DES (including, for example, differing PSF and the different galaxy redshift distributions), and that it is unlikely that our linear Taylor expansion is valid in this wide range of extracted parameter values. 
The bias in the average best-fit point over 1000 maps scales almost linearly with the masked fraction of the sky.

\begin{figure}[htb]
\includegraphics[width=3.5in]{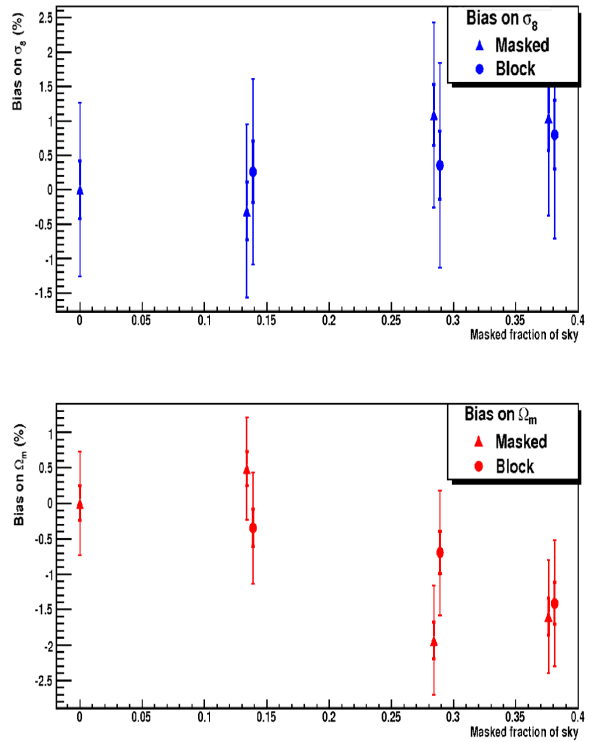}
\caption{Cosmological parameter constraints obtained if a masked dataset is compared to masked cosmological simulations. Examples of different levels of masking are shown. Points are the mean best-fit parameter estimate (offset for clarity) - triangles represent statistical uncertainties of the masked dataset, circles the uncertainties if we simply block the equivalent number of pixels. Inner error bars are the standard errors on the mean for LSST, the outer error bars are estimated for DES. }
\label{fig:bias-masked}
\end{figure}

In contrast, if we incorporate masking into our cosmological parameter estimation correctly, we see the impact of masking is greatly reduced, even with large areas of the sky masked. 
In Figure~\ref{fig:bias-masked}, we show the bias in cosmological parameter estimation with increasing levels of masking. 
We show for comparison the statistical uncertainties estimated for LSST, and estimated uncertainties for DES. 
The best-fit point does trend higher in $\sigma_8$ (and correspondingly lower in $\Omega_m$) as more masking applied. 
This is partly due to the limited size of our simulation dataset and the difference between the independent simulations used as our mock dataset, and to define the fiducial point in parameter space. 
The bias reaches 1\%\ at 37\%\ masked area, which is negligible for our estimated uncertainties for a DES-type dataset, but will be more significant for LSST.  
With a larger set of simulated lensing maps, we expect this bias to be reduced.

Figure~\ref{fig:bias-masked} also shows the parameter constraints that would be obtained if we simply blocked out the appropriate percentage of the sky. 
A comparison of the blocked and masked error bars shows that we lose less information by constructing the aperture mass statistic than we might expect if we simply removed those pixels from the map. 
For example, for a mask covering 28\%\ of the sky, the uncertainties are 10.4\%\ smaller for $\sigma_8$, and 14\%\ smaller for $\Omega_m$ compared to simply removing an equivalent number of aperture mass pixels from the survey.

\subsection{Systematic Uncertainty due to Limited Survey Area}
\label{sec:area}
In this section, we investigate the systematic uncertainty inherent in forward-modeling measurements of shear peaks statistics. 
When forward-modeling the statistics of peak counts (and indeed any statistic of weak lensing), we make the assumption that the structure in our cosmological simulations is the same as we see in the Universe. 
In our parameter estimation we are comparing peaks counts in our mock dataset to our theoretical predictions derived from averaging over many simulated maps. 
In the limit that our dataset is large, the differences in structure between the data and the theoretical predictions will be negligible because we average over many maps. 
If, however, our dataset is small, then the difference in the average structure in our data and the average structure in our simulations could be large.

To evaluate this effect, we run toy experiments with mock datasets corresponding to the size of existing and near-future lensing dataset - the DLS, CFHT-wide and DES surveys. 
For each survey, we draw the appropriate number of maps at random from our set of 1000 mock data maps to replicate the survey area, and calculate the average of the best-fit cosmological parameters over those maps. 
This is repeated 1000 times. 
The standard deviation of the scatter in the resulting set of 1000 parameter estimates is a measure of the systematic uncertainty that should be assumed when estimating cosmological parameters using the forward-modeling approach, with peak counts in maps of aperture mass. 
These are given in Table~\ref{tab:surveySyst}. 
The systematic uncertainty in $\sigma_8 (\Omega_m)$ ranges from 19\%(34\%) for a DLS-size surveys, to 11\%(20\%) for a CFHT-size survey, down to 2\%(3\%) for a DES-size survey. 
We note that we are assuming LSST-type measurement errors averaged over 100 exposures for all these three test surveys, which is certainly not the case in reality. 
We have scaled the uncertainties to account for the different $n_{\rm eff}$ of the different surveys (assuming this uncertainty simply scales as $1/\sqrt{n_{\rm eff}}$). 
The shallower depth of the DES and CFHT surveys will be an additional factor in the scatter in these numbers, as will other errors we have neglected. 
Under these caveats, this is a useful comparison of the possible utility of forward-modeling approach with different sized surveys. 
In effect, cosmic variance limits the utility of the forward-modeling approach for small survey areas.

In Table~\ref{tab:surveySyst}, we also look at the additional impact of masking on this systematic uncertainty. 
We find that even with large amounts of masking applied (a situation which we note is unlikely to occur in a real survey) the additional uncertainty due to masking is relatively small, although more significant for small survey areas.

\begin{table}
\begin{centering}
\caption[]{\textit{Percentage systematic uncertainty on parameter estimates due to differences in cosmological structure in limited datasets and simulations, for peak counts in maps of aperture mass.}}
\begin{tabular}{|l|c|c|c|c|} 
\hline
 & DLS & CFHTwide & DES \\ 
& (20 sq deg, & (154 sq deg, & (5000 sq deg, \\ 
& $n_{\rm eff}=17$) & $n_{\rm eff}=8$) & $n_{\rm eff}=13$) \\ \hline
No mask, $\sigma_8$ & 18.6\% & 11.1\% & 1.6\% \\ 
No mask, $\Omega_m$ & 33.3\% & 20.2\% & 3.0\% \\ \hline

13.4\%\ masked, $\sigma_8$ & 18.6\% & 10.7\% & 1.6\% \\ 
13.4\%\ masked, $\Omega_m$ & 33.8\% & 19.4\% & 3.0\% \\ \hline

28.4\%\ masked, $\sigma_8$ & 20.5\% & 11.4\% & 2.1\% \\ 
28.4\%\ masked, $\Omega_m$ & 36.3\% & 20.2\% & 2.9\% \\ \hline

37.6\%\ masked, $\sigma_8$ & 21.1\% & 12.4\% & 2.1\% \\ 
37.6\%\ masked, $\Omega_m$ & 36.8\% & 21.7\% & 3.0\% \\ \hline

\hline
\end{tabular}\label{tab:surveySyst}
\end{centering}
\end{table}

\section{Conclusion}
\label{sec:conclusion}

We have explored the impact of masks on shear peak counts from SNR maps of aperture mass, comparing the constraints on cosmological parameters that we might expect for unmasked and masked mock survey volumes. 
We use masks adapted from the DLS survey, and have explored up to 37.6\%\ of sky masked. 
We take a forward-modeling approach, where we combine cosmological simulations and the LSST image simulator to create mock catalogues of survey data that we might expect to see under different cosmological models. 
A comparison of the peak counts obtained from these mock surveys and the peak counts obtained from our quasi-independent mock `dataset' allows us to predict how well we will be able to constrain the cosmological parameters under consideration with an LSST-type survey. 
We find small (O(1$\sigma$) for LSST) bias in parameter estimation, even with large amounts of masking. 
This level of bias is negligible for current lensing surveys such as DES, but will become important for precision measurements with LSST. 
More cosmological simulations will be required to remove the bias caused by our limited simulation volume. 
We have also investigated the impact of differing cosmological structure in simulations and data, and have estimated the systematic uncertainty inherent in forward-modeled measurements of current survey data areas. 
We find that masking has a negligible additional impact for larger surveys, but somewhat more significant for small surveys.

The most powerful approaches to constraining cosmological parameters utilize joint analyses of many probes, for example lensing shear tomography, magnification and peak counting combined with measures of large scale-structure and supernovae. 
Such a combined approach can remove degeneracies and reduce dependence on systematic errors. 
The statistics of shear peak counts have already been shown to be a valuable constraint on cosmological parameters, especially in combination with other lensing statistics~\citep{Hilbert++2012}. 
In the absence of a reliable method of predicting shear peak counts analytically over the full range of SNR, and in the presence of survey effects such as measurement error and survey masks, the only way to constrain cosmological parameters using the statistics of shear maps is through a forward-modeling approach. 
We have shown in previous work that this approach is robust to measurement error, and in this paper we show that it is robust for survey masks and that it is a technique best suited to large survey volumes. 
Future work will concentrate on other measurement effects that must be accounted for, such as survey depth and redshift error. 
The utility of forward-modeling depends on our ability to simulate the Universe correctly, including baryonic effects and intrinsic alignments that we have neglected. 
If all these effects can all be modeled correctly in mock catalogues, then an application to survey data will be straightforward.

\section{acknowledgments}
\label{sec:acknowledgments}

The authors would like to thank Tony Tyson and Rachel Mandelbaum for their careful review of this paper on behalf of the LSST collaboration.  
The authors would also like to thank Michael Schneider for very helpful discussions, particularly related to to aspects concerning the simulations and inferred cosmological constraints.
We also thank the Deep Lens Survey for the survey masks used in this analysis, in particular Perry Gee who generated the bulk of the masks.
WD was supported in part by the NSF under Grant No. AST-1108893.
This work performed under the auspices of the U.S. Department of Energy by Lawrence Livermore National Laboratory under Contract DE-AC52-07NA27344.


\begin{thebibliography}{}
\bibitem[Ade et al(2013)]{Planck}
Ade, P. et al., 2013, 
arXiv 1303:5076. 

\bibitem[Bard et al.(2013a)]{Bard++2013}
Bard, D. et al., 2013,
Ap. J. {\bf{774}}, 49. 

\bibitem[Bard et al.(2013b)]{Bard++2013gpu}
Bard, D. et al., 2013,
As. Comp. {\bf{1}}, 17.

\bibitem[Bernstein \& Jarvis(2007)]{BernsteinJarvis2007}
Bernstein, G. M. \& Jarvis, M., 2007, 
Ap. J {bf{123}}, 2. 

\bibitem[Bartelmann \& Schneider(2001)]{BS2001}
Bartelmann, M., \& Schneider, P.,  2001, Physics Reports, {\bf 340}  291. 


\bibitem[Bertin \& Arnouts(1996)]{sEx}
Bertin, E., Arnouts, S.,  1996, AAS {\bf 117}, 393. 


\bibitem[Chang et al.(2013)]{Chang++2013}
Chang, C., et al.,  2013, 
Mon.\ Not.\ Roy.\ Astron.\ Soc.\  {\bf{47}}, 2572. 


\bibitem[Coil et al.(2004)]{coil04}
Coil, A.  et al.,  2004, Astrophys.\ J.\   {\bf 609},  525. 


\bibitem[Dietrich et al.(2007)]{Dietrich++2007}
Dietrich, J. P., et al., 2007, 
A \& A {\bf{470}}, 3. 


\bibitem[Dietrich \& Hartlap(2010)]{DietrichHartlap2010}
Dietrich, J. P. \& Hartlap, J., 2010, 
Mon.\ Not.\ Roy.\ Astron.\ Soc.\ {\bf 402}, 1049. 


\bibitem[Fan et al.(2010)]{Fan++2010}
Fan, Z., Shan, H. \& Liu, J., 2010
Ap. J. {\bf{719}}, 2.

\bibitem[Hamana et al.(2004)]{Hamana++2004}
Hamana, T., Takada, M. \& Yoshida, N.,  2004, 
  Mon.\ Not.\ Roy.\ Astron.\ Soc.\  {\bf 350}, 893. 


\bibitem[Hetterscheidt et al.(2005)]{Hetterscheidt++2005}
 Hetterscheidt, M., Erben, T. \& Schneider, P.,  2005, 
A \& A {\bf 442}, 43. 


\bibitem[Hilbert et al.(2012)]{Hilbert++2012}
Hilbert, S., Marian, L., Smith, R. E. \& Desjacques, V., 2012, 
 Mon.\ Not.\ Roy.\ Astron.\ Soc.\  {\bf 426}, 2870. 



\bibitem[Jain et al.(2000)]{Jain++2000}
Jain, B., Seljak, U. \& White, S., 2000, 
\newblock Astrophys.\ J.\   {\bf 530}, 547. 

\bibitem[Joachimi et al.(2012)]{Joachimi++2013}
Joachimi, B. et al.,  2013, 
Mon.\ Not.\ Roy.\ Astron.\ Soc.\  {\bf 431}, 477.

\bibitem[Kaiser et al.(1995)]{ksb}
Kaiser, N., Squires, G. \& Broadhurst, T.  1995, Astrohphys.\ J.\  {\bf 449},  4
60. 


\bibitem[Kratochvil et al.(2010)]{Kratochvil++2010}
  Kratochvil, J. M., Haiman, A. \& May, M., 2010, 
      Phys.\ Rev.\  D {\bf 81}, 043519. 


\bibitem[Kratochvil et al.(2012)]{Kratochvil++2012}
Kratochvil, J. M., Wang, S., Lim, E. A. et al., 2012, 
Phys. Rev. D 85, 103513. 


\bibitem[Leauthaud et al.(2007)]{Leauthaud++2007}
Leauthaud, A., Massey, R., Kneib, J-P., et al., 2007,  Astrophys.\ J.\  S {\bf 172}, 1. 



\bibitem[Lewis et al.(2000)]{Lewis++2000}
  Lewis, A., Challinor, A., \& Lasenby, A., 2000, 
  Astrophys.\ J.\  {\bf 538}, 473.

\bibitem[Liu et al.(2013)]{Liu++2013}
Liu, J. et al., 2013, 
arXiv 1310:7517. 

\bibitem[Liu et al.(2014)]{Liu++2014}
Liu, X. et al., 2014, 
Ap. J. {\bf{784}}, 1. 

\bibitem[Marian et al.(2013)]{Marian++2013}
  Marian, L., Smith, R. E., Hilbert, S. et al, 2013, 
  Mon.\ Not.\ Roy.\ Astron.\ Soc.\ {\bf 432}, 1338. 

\bibitem[Marian et al.(2009)]{Marian++2009}
Marian, L.m Smith, R. E., Bernstein, G., 2009, 
Ap. J. {\bf{698}}, 33. 


\bibitem[Maturi et al.(2011)]{Maturi++2011}
  Maturi, M., Fedeli, C., \& Moscardini, L., 2011,
  Mon.\ Not.\ Roy.\ Astron.\ Soc.\ {\bf 416}, 2527.



\bibitem[Navarro et al.(1996)]{Navarro++1996}
Navarro, J., Frenk, C. \& White, S., (1996),
  Astrophys.\ J.\ {\bf 462}, 563. 

\bibitem[Peterson et al.(in prep.)]{Peterson++prep}
Peterson, J. et al., in prep. 

\bibitem[Petri et al.(2013)]{Petri++2013}
Petri, A., et al., 2013, 
 Phys. Rev. D {\bf{88}}, 123002. 

\bibitem[Petri et al.(2014)]{Petri++2014}
Petri, A., et al., 2014, 
 arXiv:1409.5130. 

\bibitem[Saha et al.(2013)]{opsim}
Saha, A., et al., 2013, 
AAS Meeting Abstracts {\bf{221}}, 247. 

\bibitem[Schirmer et al.(2007)]{Schirmer++2007}
Schirmer, M., et al., 2007, 
A+A {\bf 462}, 875. 

\bibitem[Schneider(2005)]{Schneider2005}
Schneider. P., 2005, arXiv:astro-ph/0509252. 


\bibitem[Shirasaki et al.(2013)]{Shirasaki++2013}
Shirasaki, M., Yoshida, N. \& Hamana, T, 2013, 
Ap. J. {\bf{774}},  2.

\bibitem[Springer(2005)]{Springer2005}
  Springel, V., 2005, 
  Mon.\ Not.\ Roy.\ Astron.\ Soc.\  {\bf 364}, 1105.



\bibitem[VanderPlas et al.(2012)]{VanderPlas++2012}
VanderPlas, J., et al. 2012, Astrophys.\ J.\  {\bf 744},  180. 

\bibitem[Wittman et al.(2002)]{Wittman++2002}
Wittman, D. et al., 2002,
SPIE Proceedings {\bf{2846}}. 

\bibitem[Wittman et al.(2009)]{LSST-WL}
Wittman, D. et al.,  2009, LSST Science Book, Chapter 14, arXiv:0912.0201, http://www.lsst.org/lsst/scibook.  

\bibitem[Yang et al.(2011)]{Yang++2011}
Yang, X., Kratochvil, J.M.,  Wang, S. et al., 2011, 
  Phys.\ Rev.\  D {\bf 84}, 043529. 


\bibitem[Yang et al.(2013)]{Yang++2013}
Yang, X.,  Kratochvil, J. M.,  Huffenberger, K. M. et al., 2013, 
Phys. Rev. D {\bf 87}, 2. 


\end{thebibliography}
\end{document}